\documentclass[preprint,12pt]{elsarticle}

\usepackage{amssymb}

\journal{INTERMETALLICS}
\usepackage{graphicx}
\begin{document}

\begin{frontmatter}



\title{Half-metallic properties for the Ti${_2}$$Y$$Z$ ($Y$=Fe, Co, Ni, $Z$=Al, Ga, In) Hesuler alloys: A first-principles study}


\author{Xiao-Ping Wei}
\author{Jian-Bo Deng}
\author{Ge-Yong Mao}
\author{Shi-Bin Chu}
\author{Xian-Ru Hu}\ead{huxianru@lzu.edu.cn}

\address{Department of Physics, LanZhou University, Lanzhou 730000, People's Republic of China}

\begin{abstract}
Using the full-potential local orbital minimum-basis method, the Ti${_2}$-based full-Heusler alloys are studied. The results show that these compounds exhibit a half-metallic behavior, however, in contrast to the conventional full-Heusler alloys, the full-Heusler alloys show a Slater-Pauling behavior and the total spin magnetic moment per unit cell ($M{_t}$) following the rule $M{_t}$=$Z{_t}$-18. The origin of the gap in these half-metallic alloys are well understood. It is found that the half-metallic properties of Ti${_2}$-based compounds are insensitive to the lattice distortion and a fully spin polarization can be obtained within a wide range of lattice parameters. This is favorable in practical application.

\end{abstract}


\end{frontmatter}


\section{Introduction}
In the last decade, the half-metallic materials, exhibiting a complete spin polarization at the Fermi level, have received growing widespread attention due to its realistic applications for spintronic devices. \cite{C} Which offers opportunities for a new generation of devices combining standard microelectronics with spin-dependent effects such as nonvolatile magnetic random access memories and magnetic sensors. \cite{G} The first predicted half-metallic material was the half-Heusler alloy NiMnSb found by de Groot and collaborators in 1983. \cite{R} Since then, much attention has been paid to the Heusler alloys for new candidates. Many Heusler alloys have been predicted theoretically to be half-metals, many experiments also were carried out to establish their magnetic and transport properties in quick succession. \cite{S.P, A, S.I, Y, R.Y, S.W, J, K.R, K, Ro, A.K} \par
Heusler alloys are represented in general by the generic formula $X{_2}YZ$, where $X$ and $Y$ are transition metal elements and $Z$ is a $s$-$p$ element. \cite{P, C.C} Usually, the Heusler alloys crystallizes in a highly ordered $L2{_1}$ structure and the prototype of structure is the Cu${_2}$MnAl alloy, which belongs to the $Fm\overline{3}m$ space group. In past studies, the most Cr, Mn, Ni and Co-based Heusler alloys have been theoretically predicted to be half-metals in the $L2{_1}$ structure. \cite{Se, I.G, J.K, S, S.F, S.Is, S.Pi, K.H, R.W, C.J, K.O, K.Oz, I, M, H} Few alloys with Hg$_{2}$CuTi-type structure were investigated. \cite {R.B, G.D.L, G.D, X} Its structure is the same as that of the semi-Heusler alloys and the space group is $F\overline{4}3m$. In particular, the structure containing two Ti atoms per f.u. have rarely been studied by the electronic structure calculations up to now. In present work, our research is mainly focused on the Ti${_2}YZ$ ($Y=$Fe, Co, Ni and $Z=$Al, Ga, In) Heusler alloys with the Hg${_2}$CuTi-type structure. \par
In this paper, we present a study on a series of Ti${_2}YZ$ ($Y$=Fe, Co, Ni, $Z$=Al, Ga, In) compounds with the same structure. Firstly, we have determined that these compounds are of the Hg${_2}$CuTi-type structure instead of the conventional $L2{_1}$ structure, which will be reasoned in Sec. III. In these Ti${_2}YZ$ Heusler alloys, up to now, no reports on the half-metallicity have been found but Ti${_2}$CoAl. \cite{E} So, it is necessary to study systematically to electronic structure and magnetic properties of the Ti${_2}YZ$ ($Y$=Fe, Co, Ni, $Z$=Al, Ga, In) alloys by density functional calculations. The present paper is organized as follows. In Sec. I, the used crystal structure is illustrated. In Sec. II, we discuss the origin of the gap in these compounds. In Sec. III, the Slater-Pauling (SP) behavior of the total moments is discussed. In Sec. IV, we present the bulk and the effect of lattice constant. In Sec. V, we analysis the role of $s$-$p$ element for these Heusler alloys. Finally in Sec. VI, we summarize our results and conclusion.
\section{Calculation details}
We have carried out density functional calculations using the scalar relativistic version of the full-potential local-orbital (FPLO) minimum-basis band-structure method. \cite{K.K, I.O} In this scheme, the scalar relativistic Dirac equation was solved. For the present calculations, the site-centered potentials and densities were expanded in spherical harmonic contributions up to $\ell_{max}$ =12. For a self-consistent field iteration, the charge density is converged to 10$^{-6}$, the convergence of the total energies (10$^{-8}$ hartree) with respect to $\kappa$-space integrations was checked for each of the considered Heusler alloys. The Perdew-Burke-Ernzerhof 96 of the generalized gradient approximation (GGA) was used for exchange-correlation (XC) potential. \cite{J.P}
The self-consistent potentials were carried out on a $\kappa$ mesh of 20 $\kappa$-points in each direction of the Brillouin zone, the Brillouin-zone integrations were performed with the tetrahedron method. \par

\section{Results and discussion}
\subsection{Crystal structure}
In general, the cubic $X_{2}YZ$ Heusler alloys are found with Cu${_2}$MnAl-type structure but also with the Hg${_2}$CuTi-type structure. The Hg${_2}$CuTi-type structure exhibits $T{_d}$ symmetry. In the type of structure, the nonequivalent 4a and 4c Wyckoff position is occupied by the two $X$ atoms, the $Y$ and $Z$ are located on 4b and 4d positions, respectively. This structure is similar to the $XYZ$ alloys with $C1{_b}$ structure, the difference is only the vacancy filled by an additional $X$ atom in $T{_d}$ symmetry. It is reported if the nuclear charge of the $Y$ element is larger than the one of the $X$ element from the same period, this structure will be frequently observed. Furthermore, the structure may also appear in compounds comprised of transition metals from different periods. \cite{He} In fact, the two structures may be hardly distinguishable by x-ray diffraction and much care has to be taken in the structural analysis, as both have the general fcc-like symmetry. \par
For the Ti${_2}$$Y$$Z$ ($Y$=Fe, Co, Ni, $Z$=Al, Ga, In) Heusler alloys, it might be of interest to pay close attention to those compounds beacuse of these Ti$_{2}$-based compounds were not reported in any experiment. Within the FPLO scheme, a structural optimization was performed for these compounds to find the lattice parameter resulting in the minimum of the total energy. It was also confirmed that the ferromagnetic configuration is lower in energy than the non-spin polarized case for these compounds. An anti-parallel spin arrangement of the $sp$ atom with respect to the $X$ atoms in the cubic lattice is showed, we also observed a similar conclusion as mentioned above. \cite{X.P} Based on the conclusion, it is easy to estimate the superlattice structure of the Heusler compounds. In addition, the selectivity of occupation sites for our materials is also supported. Thus, in this paper the Hg${_2}$CuTi-type structure is applied. \par
\subsection{Origin of the gap}
In this section, we will focus on the energy gap in minority-spin states. Such a gap is the most important feature of a half-metal. In these Ti${_2}$-based alloys, the calculated results show a band gap for the minority carriers, as shown in figure 1. The origin of the band gap is usually distinguished into three categories: (1) covalent band gap, (2) $d$-$d$ band gap, and (3) charge transfer band gap. \cite{C.M} The covalent band gap has been shown to exist in semi-Heusler alloys with $C1{_b}$ structure. The $d$-$d$ band gap is responsible for the half-metallicity of the full-Heusler alloys with $L2{_1}$ structure. The charge transfer band gap is usually common in CrO${_2}$ and double perovskites. \cite{G.D}\par
In our Ti${_2}$-based alloys, the situation is rather complicated due to the special crystallized structure. The structure of Ti${_2}$-based alloys has a space group similar to the semi-Heusler alloys but the empty site is occupied. Compared with the full-Heusler alloys with $L2{_1}$ structure, the Ti atoms occupy two sublattices with different surroundings. Therefore, it is important to pay attention to the origin of band gap with Hg${_2}$CuTi-type structure.\par
To understand the origin of the band gap, let us observe the density of states (DOS) shown in figure 1. It can be seen that the Ti${_2}$-based alloys with the same $X$ and $Y$ atoms have the similar feature of DOS. The minority-spin DOS can be characterized by the large gap at the Fermi level and the occupied bonding states mainly present $Y$ atom characters below the Fermi level. It is evident that the $d$-$d$ orbitals hybridization between transition metals is rather intensive. Since the current studied Heusler alloys have a tetrahedral $T{_d}$ symmetry which is a subgroup of $O{_h}$, the possible hybridization between $d$-$d$ orbitals sitting at the different sites will be created into several energy level, such as $e{_g}$, $t_{2g}$, $e_{u}$ and $t_{1u}$. Their complex exchange splitting will shift the Fermi level to the appropriate position and is responsible for the origin of gap. Furthermore,
we can also see that the bonding hybrids are mainly localized at the high-valent transition metal atom such as Fe, Co or Ni, while the unoccupied antibonding states are mainly at the low-valent transition metal such as Ti. The bonding states preferentially reside at $Y(C)$ sites and produce the spin density of occupied states. So, the covalent hybridization between high-valent and low-valent atoms also leads to the origin of gap. \cite{Ji}\par
 As elucidated above. In these Ti${_2}$-based alloys, the states near the Fermi level can be well ascribed to covalent hybridization and $d$-$d$ orbitals hybridization between transition metals. In addition, as can be seen from the DOS patterns, the $Z$ atom plays an important role in the half-metallicity of the Heusler alloys, although it does not directly form the minority gap, which provides $s$ and $p$ states to hybridize with $d$ electrons and determines the degree of occupation of the $p$-$d$ orbital. Thus, the hybridization between $p$-$d$ electrons affects the formation of the energy gap and the width of gap.
In conclusion, it should be emphasized that there exist two mechanisms in the formation of the band gap, that is covalent band gap and $d$-$d$ band gap, but it is mainly the $d$-$d$ band gap that characterizes the half-metallicity in Ti${_2}$-based alloys.\par

\section{Magnetic properties and Slater-Pauling rule}
The preceding discussion is concerned with the electronic structure and the formation of the band gap. Now, we turn to the Slater-Pauling rule, which is a simple way to study for ferromagnetic alloys the interrelation between the valence electron concentration and the magnetic moments (see figure 2). It is well known that the $C1{_b}$ Heusler alloys with three atoms per unit cell follow a simple rule that scales linearly with the number of valence electrons: $M_{t}$ = $Z_{t}$$-$18, where $M_{t}$ represents the total spin magnetic moment and $Z_{t}$ stands for the total number of valence electrons per unit cell, which was first noted by K\"{u}bler $et$ $al$ \cite{Ku} and the importance of which was also emphasized by Jung $et$ $al$ \cite{Ju} and Galanakis $et$ $al$. \cite{I.Ga} Since with 9 electron states occupied in the minority band for $C1{_b}$ Heusler alloys. $Z_{t}$$-$18 is just the number of uncompensated electron spins. Namely, that is magnetic moment per unit cell.\par
For ordered alloys with different kinds of atoms, it might be more convenient to work with all atoms of the unit cell. In the case of four atoms per unit cell, as in Heusler alloys like Co${_2}$MnGe with $L2_{1}$ structure, one has to subtract 24 from the accumulated number of valence electrons to find the magnetic moment per unit cell ($M_{t}$):
$M_{t}$ = $Z_{t}$$-$24 \cite{G.H} with $Z_{t}$ denoting the accumulated number of valence electrons in the unit cell containing four atoms.
In the case of Heusler alloys, the number 24 arises from the number of completely occupied minority bands that has to be 12 in the half-metallic state. Particularly, these are one $s$ ($a_{1g}$), three $p$ ($t_{1u}$), and eight $d$ bands. The latter consist of two triply degenerate bands with $t_{2g}$ symmetry and one with $e_{g}$ symmetry.\par
In figure 3 we have gathered the calculated spin magnetic moments per unit cell for the these Heusler alloys which we have plotted as a function of the total number of valence electrons. The solid line represents the rule $M_{t}$=$Z_{t}$$-$18 obeyed by these Heusler alloys. The total magnetic moment $M_{t}$ (in $\mu_{B}$) is just the difference between the number of occupied spin-up states and occupied spin-down states. The number of spin-down bands below the gap is in all cases $N_{\downarrow}$=9. Thus, we can directly deduce the number $N_{\uparrow}$ = $Z{_t}$$-$$N_{\downarrow}$ = $Z{_t}$$-$9 of occupied spin-up states and the moment, $M_{t}$ = ($N_{\uparrow}$$-$$N_{\downarrow}$) $\mu_{B}$ = ($Z_{t}$$-$2$\times$$N_{\downarrow}$) $\mu_{B}$ = ($Z_{t}$$-$18) $\mu_{B}$ from the total number of valence electrons ($Z_{t}$), is given by the difference. It might be with nine minority bands fully occupied. So, we obtain the simple the rule for half-metallicity Hg${_2}$CuTi-type Heusler alloys:
$M_{t}$ = $Z_{t}$$-$18. The calculated total moment $M_{t}$ is an integer 1, 2 and 3 $\mu_{B}$ with respect to total valence electrons 19, 20, 21. Which is in good agreement with our expected results. For all these studied Heusler alloys, we find that the total spin moments scales accurately with the total valence electrons and that they all present half-metallicity.\par
\section{Bulk and Effect of the lattice parameter}
\subsection{The changes of gap with lattice constant }
In order to expound the influence of the lattice constant for width gap. In figure 3, we give the width of the minority gap with increasing the lattice parameter. Clearly, it can be seen that the increase of the lattice enlarges the energy gap for Ti${_2}$Fe-based alloys, while the width gap of Ti${_2}$Ni-based alloys is continuously decreasing with the expansion of lattice parameter. The change of the width of the gap can be traced back to two combined effects: (1) $p$-$d$ orbital hybridization, which obviously depends on the $p$ states of $s$-$p$ element and $d$ states of $Y$ atom. From Table 1, we can observe that the width of the energy gap in Ti$_{2}$Co-based alloys decreases with increasing atomic number of $s$-$p$ element. (2) their different lattice parameters. From the partial DOS of atoms, the binding energy of $p$ states of $Z$ atoms decreases generally with increasing atomic number, which seems well to illustrate the change of the gap as stated for Ti$_{2}$Co-based alloys. In fact, the effect of $p$-$d$ hybridizations covers up the effect of lattice parameter, the contraction or expansion of the lattice parameter has an influence on the delocalized $p$ electrons but not well localized $d$ electrons of the transition metal. It is likely that the decreasing binding energy squeezes the energy gap. Furthermore, we also show bulk modulus and the range of pressure existing in the half-metallicity in Table 1. Changing the lattice parameter results in shifting the Fermi level without destroying the gap. The values of the bulk modulus and range of pressure with the optimized lattice parameters for these alloys are given in Table 1. The results show that the Ti$_{2}$FeIn alloy is difficult to be compressed and bulk modulus value up to 264 GPa. The detailed values of the calculated total and atomic resolved moments are also included in Table 1.\par

\section{The influence of the main element}
In this Section, we will study on the role of the $s$-$p$ element in the electronic properties of Hg${_2}$CuTi-type structure alloys. Firstly, we focus on the influence due to atomic number change of $s$-$p$ element. Secondly, We consider further the influence of $s$-$p$ element substituted by the empty site, where we performed calculations using the same lattice parameter as $X{_2}YZ$. In short, they are very important for the physical properties of the Heusler alloys and the structural stability of the Hg${_2}$CuTi-type Heusler alloys, as we will discuss in the following. There are three important features:\par
(i) In the Ti${_2}$-based Heusler alloys, the $s$- and $p$-states hybridize with the $d$-states of transition metal. In figure 1, we can see clearly that the total DOS have almost similar feature with the change of $s$-$p$ element. In addition, the $p$ electrons of the $s$-$p$ element hybridize with the $d$ states, which determine the degree of occupation of the $p$-$d$ orbitals. We should note that the energy of the $p$ electrons is strongly dependent on $Z$ element, which can be clearly seen in the resolved DOS patterns. The hybridization between $p$ electrons with different energies and $d$ electrons is crucial to the width of the energy gap in the minority-spin band. Even though the Fermi level is shifted slightly, which remains to be pinned in the gap of minority-spin states. Therefore, we predict that partly substituting one kind of $s$-$p$ element for another might change the other properties of the material but not destroy their half-metallicity. In fact, the change of the $s$-$p$ element results in a rigid shift of the bands. \cite{I.Gal} \par

(ii) The $s$-$p$ element is very important for the structural stability, which can be seen obviously from the calculated DOS shown in figure 4, the DOS with $s$-$p$ element shift slightly to bonding states improving the structural stability and is delocalized due to the strong exchange interaction, since metallic alloys prefer highly coordinated structures like fcc, bcc, hcp, etc. Therefore, the $s$-$p$ elements are decisive for the structural stability of the Hg${_2}$CuTi-type alloys. A careful discussion of the bonding in the conventional Heusler alloys has been recently published by Nanda and Dasgupta \cite {N} using the crystal orbital Hamiltonian population (COHP) method. \par

\section{Summary and conclusions}
In summary, we have studied in detail the electronic and magnetic properties of the Ti$_{2}$-based series Heusler alloys. It is shown that the gap of the Hg$_{2}$CuTi-type alloys is imposed by the lattice structure. The hybridization between covalent bonding in high-valence and lower-valence transition metal and $d$-$d$ orbitals leads to the origin of the minority gap. \par

 The calculated total magnetic moments $M_{t}$ are 1$\mu_{B}$/f.u. for Ti$_{2}$Fe$Z$ ($Z=$Al, Ga, In), 2$\mu_{B}$/f.u. for Ti$_{2}$Co$Z$ ($Z=$Al, Ga, In), and 3$\mu_{B}$/f.u. for Ti$_{2}$Ni$Z$ ($Z=$Al, Ga, In) following the $M{_t}$=$Z{_t}$$-$18 rule, which agree with the Slater-Pauling curve quite well. The absence of an obvious variation in $M_{t}$ for different alloys is mainly ascribed to the counterbalance of the moments of $X$ and $Y$ in value and direction. The spin polarization of these alloys is found to be 100\% within the pressure range, which is attractive in practical applications. \par

According to the influence of $s$-$p$ element described, it is confirmed that the stability of the structure and the half-metallicity of the Ti$_{2}$$YZ$ ($Y=$Fe, Co, Ni and $Z=$Al, Ga, In) Heusler alloys are dominated by the presence of $s$-$p$ element. In addition, for all Ti$_{2}$-based Heusler alloys, the localized magnetic moment carried by the Ti and $Y$ atoms is restricted by the $s$-$p$ element even though they have a small contribution to the total spin magnetic moment. Further, it is found that the minority gap of Ti$_{2}$Ni$Z$ ($Z=$Al, Ga, In) alloys shows a trend to decrease with increasing lattice parameter. The reason may be that the hybridization become weaker with the expansion of lattice parameter. \par

In conclusion, $ab$ $initio$ calculations have been used to study the electronic structure and magnetic properties of state-of-the-art Heusler alloys and predict successfully Ti$_{2}$$YZ$($Y=$Fe, Co, Ni and $Z=$Al, Ga, In) to be new half-metallicity. In Hg$_{2}$CuTi-type structure, the special crystallized structure leads to a complex magnetic interaction and brings about many physical properties for the Ti$_{2}$$YZ$($Y=$Fe, Co, Ni and $Z=$Al, Ga, In) alloys. Thus, our findings suggest that these compounds can be viewed as a choice for further experimental investigations.

\section{acknowledgments}
The authors would thank to be grateful for Manuel Richter, Klaus Koepernik, Ulrike Nitzsche and Wen-Xu Zhang for many useful discussions.\par

\newpage
\begin{figure}
 \centering
  \includegraphics[width=10cm,angle=-90]{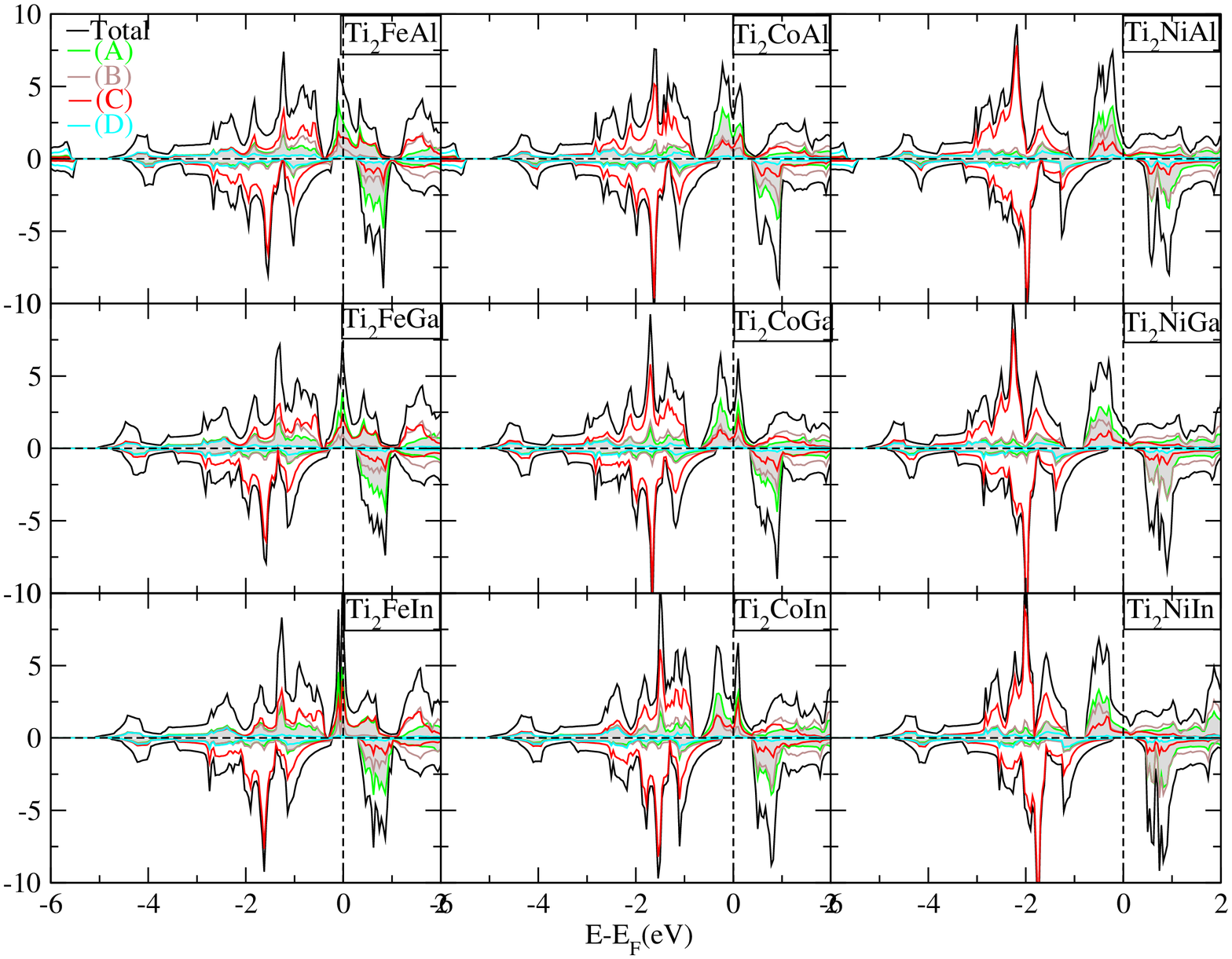}\\

\end{figure}

\begin{figure}
  \centering
  \includegraphics[width=10cm,angle=-90]{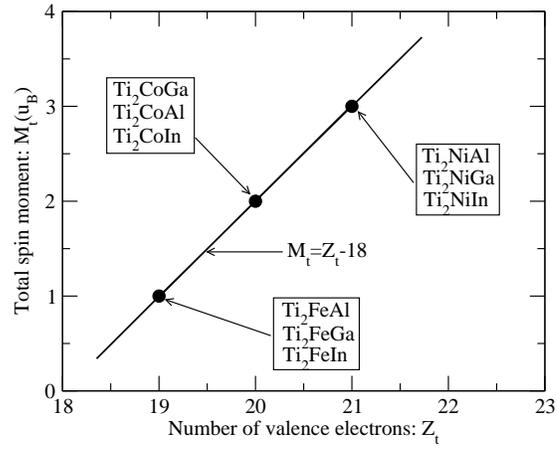}\\
  \caption{Calculated total spin moments for the studied Ti$_{2}YZ$ ($Y=$ Fe, Co, Ni and $Z=$ Al, Ga, In) alloys. The solid line represents the Slater-Pauling curve for these Heusler alloys}\label{figure}
\end{figure}

\begin{figure}
   \centering
  \includegraphics[width=10cm,angle=-90]{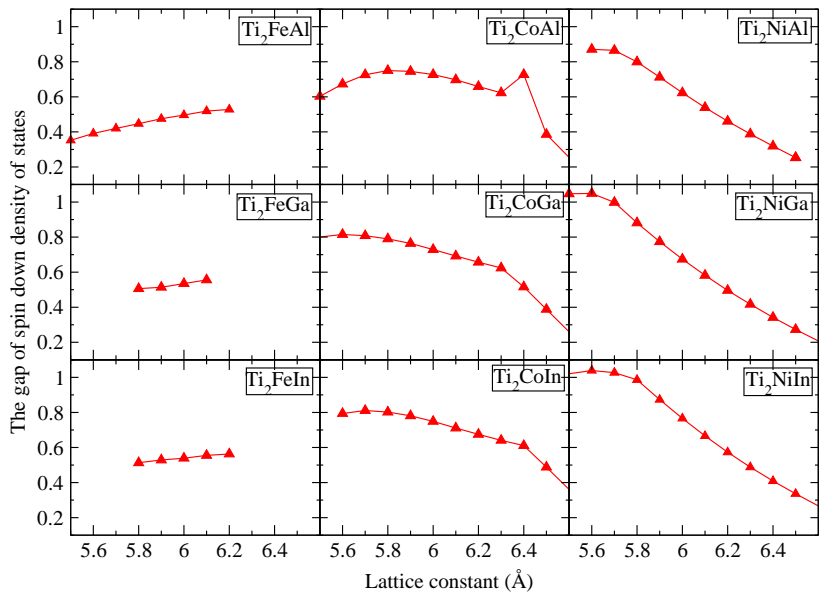}\\
  \caption{The width of minority gap changes with the lattice parameters in the range of 5.50$-$6.60${\AA}$.}\label{figure}
\end{figure}

\begin{figure}
   \centering
  \includegraphics[width=10cm,angle=-90]{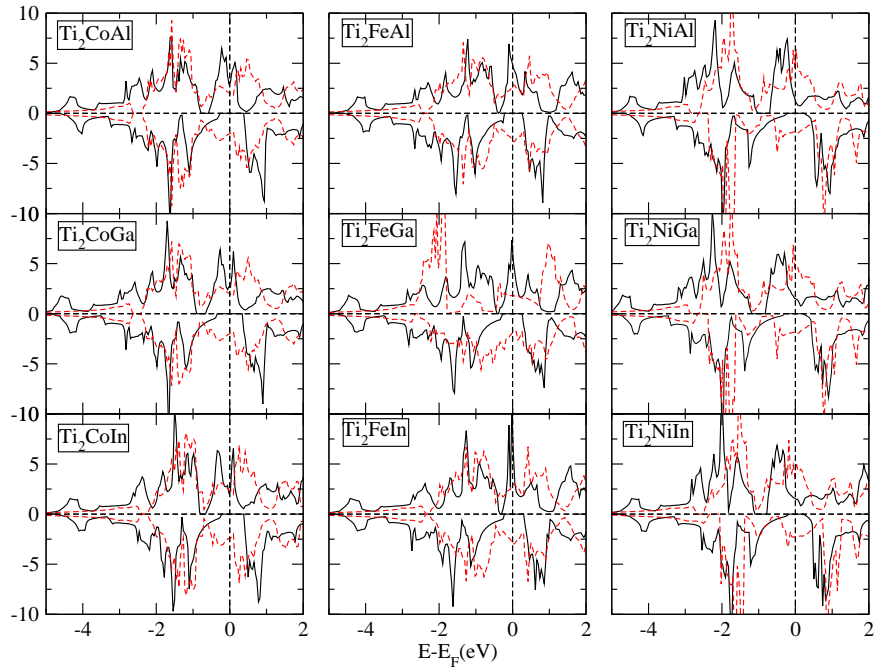}\\
  \caption{The calculated total DOS for Ti$_{2}YZ$ and Ti$_{2}Y$ alloys. Of course we should keep in mind that all these calculations have been performed at the Ti$_{2}YZ$ lattice constants. In Ti$_{2}Y$ alloys, the $Z$ atom has been subsituted with a vacant site. Solid black line represents the total DOS of Ti$_{2}YZ$ alloys, and the total DOS of Ti$_{2}Y$ alloys are represented by red dotted line }\label{figure}
\end{figure}

\begin{table}
\caption{Magnetic data, bulk modulus and width of minority gap for series Ti$_{2}YZ$ ($Y=$ Fe, Co, Ni and $Z=$ Al, Ga, In) with optimized lattice parameter, and the range of pressure keeps the half-metallic properties is also listed.}{\label{Table 1}}
\begin{tabular}{cccccccccc}
  \hline
Compounds & a$_{opt}$ & $X_{A}$ & $X_{B}$ & $Y_{C}$ & $Z_{D}$ & $M_{tot}$ & P& $B$ & gap width\\
           & (${\AA}$) & ($\mu{_B}$) & ($\mu{_B}$) & ($\mu{_B}$) & ($\mu{_B}$) & ($\mu{_B}$/f.u) & (GPa)& (GPa) & (eV)\\
  \hline
  Ti$_{2}$FeAl & 6.14 & 1.21 & 0.84 &-1.02 & -0.02 & 1.00 & 973.97  & 130.6 &0.53\\
  Ti$_{2}$FeGa & 6.12 & 1.22 & 0.93 &-1.09 & -0.06 & 1.00 & 401.50  & 125.6 &0.56\\
  Ti$_{2}$FeIn & 6.23 & 1.32 & 1.06 &-1.29 & -0.08 & 1.00 & 780.12  & 264.0 &0.56\\
  Ti$_{2}$CoAl & 6.14 & 1.50 & 0.80 &-0.23 & -0.07 & 2.00 & 899.55  & 137.4 &0.68\\
  Ti$_{2}$CoGa & 6.12 & 1.45 & 0.88 &-0.24 & -0.09 & 2.00 & 1197.55 & 78.8  &0.68\\
  Ti$_{2}$CoIn & 6.36 & 1.54 & 1.00 &-0.42 & -0.12 & 2.00 & 1209.13 & 129.3 &0.62\\
  Ti$_{2}$NiAl & 6.20 & 1.83 & 1.15 & 0.10 & -0.08 & 3.00 & 892.56  & 127.5 &0.46\\
  Ti$_{2}$NiGa & 6.19 & 1.78 & 1.24 & 0.10 & -0.11 & 3.00 & 962.82  & 125.3 &0.50\\
  Ti$_{2}$NiIn & 6.42 & 1.79 & 1.28 & 0.07 & -0.15 & 3.00 & 1208.90 & 119.6 &0.39\\
  \hline
\end{tabular}
\end{table}

\end{document}